\documentclass[10pt,letterpaper]{article}

\usepackage{opex3}
\usepackage{amsmath,amsfonts}
\usepackage[amssymb]{SIunits}

\renewcommand{\vector}[1]{\mathbf{#1}}

\renewenvironment{abstract}
{\vskip1pc\noindent\begin{center} \begin{minipage}{.8\textwidth} {\bf Abstract: \ } }
{ \\ \vskip-.5pc \noindent \small \copyright \, 2009 \hskip.05in
   Optical Society of America \\ \hfil \end{minipage}\end{center}\normalsize\vskip-1.5pc}%

\begin{document}

\title{Analytical solution for wave propagation through a graded index interface between a right-handed and a left-handed material}

\author{Mariana~Dalarsson$^1$ and Philippe~Tassin$^2$}
\address{$^1$Royal Institute of Technology, Falkv 7, 125 34 Alvsjo - Stockholm, Sweden}
\email{mardal@kth.se}
\address{$^2$Dept.~of~Applied~Physics~and~Photonics, Vrije~Universiteit~Brussel,\\
Pleinlaan~2, B-1050~Brussels, Belgium}
\email{philippe.tassin@vub.ac.be}

\begin{abstract}
We have investigated the transmission and reflection properties of structures incorporating left-handed materials with graded index of refraction. We present an exact analytical solution to Helmholtz' equation for a graded index profile changing according to a hyperbolic tangent function along the propagation direction. We derive expressions for the field intensity along the graded index structure, and we show excellent agreement between the analytical solution and the corresponding results obtained by accurate numerical simulations. Our model straightforwardly allows for arbitrary spectral dispersion.
\end{abstract}


\bibliographystyle{osajnl}

\section{Introduction}

During the last decade, a new class of artificial composite materials called electromagnetic metamaterials has emerged. Such metamaterials are structured at a subwavelength level and have electromagnetic resonators or ``particles'' such as split-ring resonators and nanowires as their structural units~\cite{Soukoulis-2006}. Through the appropriate design of the electric and magnetic resonances of these subwavelength particles, it is possible to create materials with electromagnetic properties that are unknown for conventional materials. Particularly, there has been considerable interest in metamaterials with simultaneously negative permittivity and negative permeability, which are called left-handed materials. It was shown in the theoretical work of Veselago~\cite{Veselago} that left-handed materials exhibit a number of remarkable properties, including negative index of refraction (and, hence, negative phase velocity), inverse Doppler effect, and radiation tension instead of pressure. All these properties stem from the fact that the Poynting vector in these materials is antiparallel to the wavevector, i.e., the electric field, the magnetic field and the wavevector of a plane electromagnetic wave form a left-handed system of reference.

The concept of metamaterials was brought to practical implementation by a number of works by Pendry~\cite{Pendry1,Pendry2}, who suggested split-ring resonators and wire arrays as the first metamaterial particles, and the first experimental demonstration of a left-handed material was published in 2001~\cite{Shelby-2001}. Split-ring resonators and nanowires are still widely used in the microwave domain and are now well understood, but many other particles such as slab-wire pairs, fishnets~\cite{Soukoulis-2006} and coupled split-ring resonators~\cite{Tassin-2009} have been designed. One of the main challenges in the field of optical metamaterials is the reduction of the relatively high losses associated with their resonant behavior. The development of metamaterials with permittivity and/or permeability negative or smaller than unity has also led to a revolution in our understanding of optics; several remarkable applications have been proposed, including superlenses and hyperlenses that enable imaging below the diffraction limit~\cite{Pendry3,Fang,Jacob}, waveguides that can stop light~\cite{Tsakmakidis-2007}, miniaturized photonic devices such as Fabry-perot resonators and waveguides~\cite{Engheta-2007,Tassin-2008}, and even invisibility cloaks through the technique of transformation optics~\cite{Leonhardt-2006,Pendry-2006}.

In this paper, we want to consider the transmission and reflection properties of graded index optical structures with a gradual transition from a right-handed to a left-handed material. Such structures were studied in the framework of metamaterial gradient index lenses by a few authors~\cite{Ramakrishna2,Pinchuk,Parazzoli,Dalarsson}, who have shown that this provides an additional degree of freedom that can be used, among others, to reduce geometrical aberrations; a gradient index metamaterial lens was also demonstrated experimentally by Smith~\cite{Smith2}. These works addressed the propagation through graded index structures with geometrical optics and therefore neglect any reflections that may occur on such structures. This issue has been taken up only very recently~\cite{Litchinitser}.

\begin{figure}[th!]
  \centering
  \includegraphics[width=5.5cm]{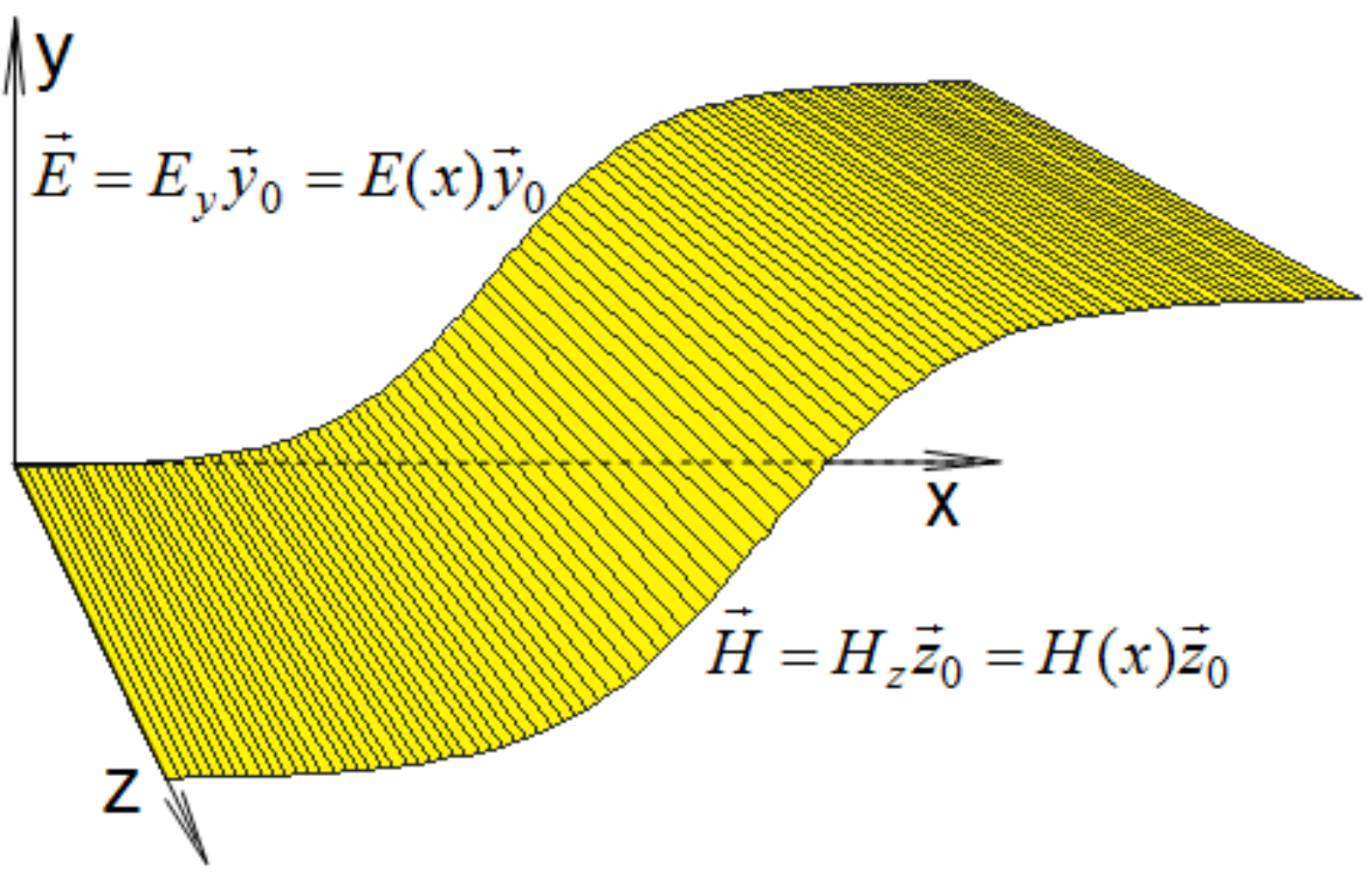}
  \caption{Propagation of an electromagnetic wave through a graded index structure with a hyperbolic tangent profile. This is the index profile assumed in this paper.}
  \label{Fig:Figure1}
\end{figure}

Here we present an exact analytical solution of Helmholtz' equation for the propagation of electromagnetic waves through a graded index metamaterial structure. We choose a graded index profile for which both the permittivity and the permeability vary according to a hyperbolic tangent function (see Fig.~\ref{Fig:Figure1}). In Sec.~\ref{Sec:FieldEquations}, we review the field equations that the electric and magnetic fields must satisfy in an inhomogeneous medium and we transform these equations to simple wave equations with an effective wavenumber. In Sec.~\ref{Sec:Analytical}, we present the analytical solution for the waves propagating in this structure. Finally, we compare the obtained analytical solution with results obtained from accurate numerical simulations based on a finite element method in Sec.~\ref{Sec:Numerical}.

\section{Field equations}\label{Sec:FieldEquations}

We start our analysis from Maxwell's equations, and we search for fields that are periodic in time according to a $\exp(-i \omega t)$ dependency. Furthermore, we assume that the effective medium approximation can be made and that the materials are isotropic, so that their optical properties can be described by the effective dielectric permittivity and the effective magnetic permeability. For most metamaterials, the effective medium assumption is valid, because their constituents elements are on the subwavelength level. The geometry of the problem is illustrated in Fig.~\ref{Fig:Figure1}.
The electric field is directed along the $y$-axis, $\vector{E}(\vector{r}) = E(x) \vector{e}_\mathrm{y}$, whereas the magnetic field is directed along the $z$-axis, $\vector{H}(\vector{r}) = H(x) \vector{e}_\mathrm{z}$. The propagation direction of the wave is along the $x$-axis. Since the fields depend only on the $x$-coordinate, we have 
\begin{align} 
\frac{dE}{dx} &= i \omega \mu H(x),\label{Eq1}\\
\frac{dH}{dx} &= i \omega \epsilon E(x),\label{Eq2}
\end{align} 
where $\epsilon = \epsilon(\omega, x)$ and $\mu = \mu(\omega, x)$ are the frequency-dependent dielectric permittivity and magnetic permeability, respectively.

We can easily eliminate either the electric field or the magnetic field from Eqs.~(\ref{Eq1})-(\ref{Eq2})~\cite{Yeh}; this yields an ordinary differential equation for either 
$E(x)$ or $H(x)$:
\begin{align} 
&\frac{d^2E}{dx^2} - \frac{1}{\mu} \frac{d\mu}{dx} \frac{dE}{dx} + \omega^2 \mu \epsilon E(x) = 0,\label{Eq3}\\
\intertext{or}
&\frac{d^2H}{dx^2} - \frac{1}{\epsilon} \frac{d\epsilon}{dx} \frac{dH}{dx} + \omega^2 \mu \epsilon H(x) = 0.\label{Eq4}
\end{align}
These equations describe the propagation of electromagnetic waves through a medium of which the constitutive parameters vary along the propagation direction. The spatial dependency of the functions $\epsilon(x)$ and $\mu(x)$ may be completely arbitrary, even on space scales faster than the wavelength of the radiation, on the condition of course that the effective medium approximation remains valid. 

The standard approach to the solution of Eqs.~(\ref{Eq3})-(\ref{Eq4}) is to eliminate the first order terms by introducing the functions $F(x)$ and $G(x)$ instead of the functions $E(x)$ and $H(x)$ using the following transformations:
\begin{equation} 
E(x) = \sqrt{\mu(x)} F(x),\hspace{5mm} H(x) = \sqrt{\epsilon(x)} G(x).
\label{Eq5}
\end{equation} 
In this way, we obtain the following wave equations for the functions $F(x)$ and $G(x)$:
\begin{align} 
\frac{d^2F}{dx^2} + \left[ \omega^2 \mu \epsilon + \frac{1}{2 \mu} \frac{d^2\mu}{dx^2} -
\frac{3}{4 \mu^2} \left( \frac{d\mu}{dx} \right)^2\right] F(x) &= 0, 
\label{Eq6}\\
\intertext{and}
\frac{d^2G}{dx^2} + \left[ \omega^2 \mu \epsilon + \frac{1}{2 \epsilon} \frac{d^2\epsilon}{dx^2}
- \frac{3}{4 \epsilon^2} \left( \frac{d\epsilon}{dx} \right)^2\right] G(x) &= 0.\label{Eq7}
\end{align}
These two equations can also be written as wave equations
\begin{equation} 
\frac{d^2F}{dx^2} + k^2_{\mu}(x) F(x) = 0, \hspace{5mm}\frac{d^2G}{dx^2} + k^2_{\epsilon}(x) G(x) = 0, \label{Eq8}
\end{equation}
where 
\begin{align} 
&k^2_{\mu}(x) = \omega^2 \mu \epsilon + \frac{1}{2 \mu} \frac{d^2\mu}{dx^2} -
\frac{3}{4 \mu^2} ( \frac{d\mu}{dx} )^2\label{Eq9}\\
\intertext{and}
&k^2_{\epsilon}(x) = \omega^2 \mu \epsilon + \frac{1}{2 \epsilon} \frac{d^2\epsilon}{dx^2}
- \frac{3}{4 \epsilon^2} ( \frac{d\epsilon}{dx} )^2\label{Eq10}
\end{align}
are the space-dependent effective wave vectors for electric and magnetic fields respectively. In case of the hyperbolic tangent profile for the functions $\epsilon(x)$ and $\mu(x)$, Eqs.~(\ref{Eq6})-(\ref{Eq7}) are generally reduced to the hypergeometric equation, allowing for analytical solution in terms of suitable hypergeometric functions.       

\section{Analytical solutions of the field equations}\label{Sec:Analytical}
\begin{figure}[tb]
  \centering
  \includegraphics[width=7cm]{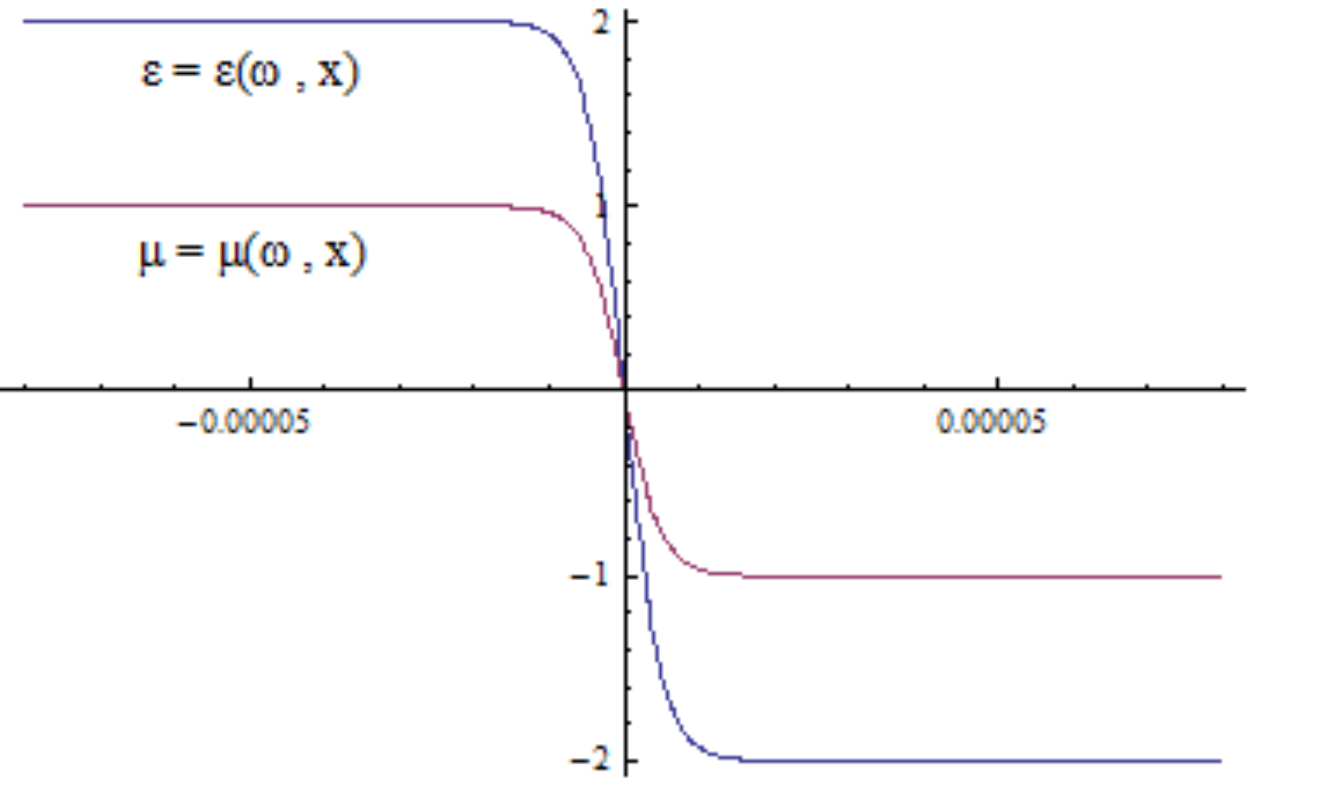}
  \caption{We assume in this paper that the effective permittivity $\epsilon_{eff}$ and permeability $\mu_{eff}$ vary along the propagation direction according to a hyperbolic tangent function.}
  \label{Fig:Figure2}
\end{figure}

In this paper, we consider an inhomogeneous medium for which the effective permittivity and permeability vary according to a hyperbolic tangent function. We choose the hyperbolic tangent function here because we will show that it is possible to construct analytical solutions for this case. Furthermore, it is a convenient function that naturally provides asymptotic values of the constitutive parameters in the backward and forward directions and allows a detailed study of the limit of the abrupt transition as well. We use the antisymmetric functions (see Fig.~\ref{Fig:Figure2})
\begin{equation} 
\mu = - \mu_0 \mu_{eff}(\omega) \tanh (\rho x), \hspace{5mm}
\epsilon = - \epsilon_0 \epsilon_{eff}(\omega) \tanh (\rho x),
\label{Eq11}
\end{equation}
where $\rho$ is a positive real parameter describing the steepness of the transition from the right-handed material at the left-hand side of the plane $x = 0$ to the left-handed material at the right-hand side of the plane $x = 0$. There is no restriction on the functions $\mu_{eff}(\omega)$ and $\epsilon_{eff}(\omega)$ (except of course for such restrictions as the Kramers-Kronig relationships), so that our method allows for arbitrary spectral dispersion. The reader should note that the impedance $Z = Z_0 Z(\omega) = \sqrt{\mu_0 \mu_{eff}(\omega)/\epsilon_0 \epsilon_{eff}(\omega)}$ is constant throughout the entire structure; as a result, there is no reflection on the graded interface between the two materials.

We found that, for this particular graded index structure, the two differential equations~(\ref{Eq3})-(\ref{Eq4}) have a remarkably simple set of two independent exact solutions given by
\begin{equation}
E(x) = E_0 \left[ \cosh \left(\rho x\right) \right]^{\pm i \frac{\kappa}{\rho}}, \hspace{5mm}
H(x) = H_0 \left[ \cosh \left(\rho x\right) \right]^{\pm i \frac{\kappa}{\rho}},
\label{Eq13}
\end{equation}
where $E_0$ and $H_0$ are constant amplitudes, and 
\begin{equation}
\kappa^2 = \omega^2 \epsilon_{eff}(\omega) \mu_{eff}(\omega). 
\label{Eq14}
\end{equation}
We repeat that these exact solutions are valid for arbitrary steepness $\rho$. Let us now choose the solution with the minus sign in the exponent of the expression~(\ref{Eq13}), i.e., 
\begin{equation}
E(x) = E_0 \left[ \cosh \left(\rho x\right) \right]^{-i \frac{\kappa}{\rho}}.\label{Eq15}
\end{equation}
From the asymptotic expansion of the field $E(x)$ in the limits $x\rightarrow \mp \infty$,        
\begin{equation} 
E(\mp\infty) = E_0 e^{ i \frac{\kappa}{\rho} \ln 2} e^{\pm i \kappa x},\label{Eq16}
\end{equation}
we see that the wave described by Eq.~(\ref{Eq15}) is an electromagnetic wave with wavevector 
$\vector{k_{-\infty}}~=~+\kappa\vector{e}_\mathrm{x}$ in the right-handed material far from the interface ($x\rightarrow -\infty$). This is a wave that propagates in the $+x$ direction, i.e., a wave propagating to the right. On the other hand, for $x\rightarrow +\infty$, the wave has wavevector $\vector{k_{+\infty}} = - \kappa\vector{e}_\mathrm{x}$; this represents a wave of which the phase fronts propagate in the $-x$ direction. However, since we have a left-handed material for $x>0$, the energy flux (Poynting's vector) is still propagating from left to right. This is perfectly consistent with the fact that there is no reflection on this structure. This is also apparent from the fact that $|\cosh(\rho x)^{-i\kappa/\rho}| = 1$, so that $|E(x)|$ is constant throughout the structure.

\section{Comparison with numerical results}\label{Sec:Numerical}
\begin{figure}[tp]
  \centering
  \includegraphics{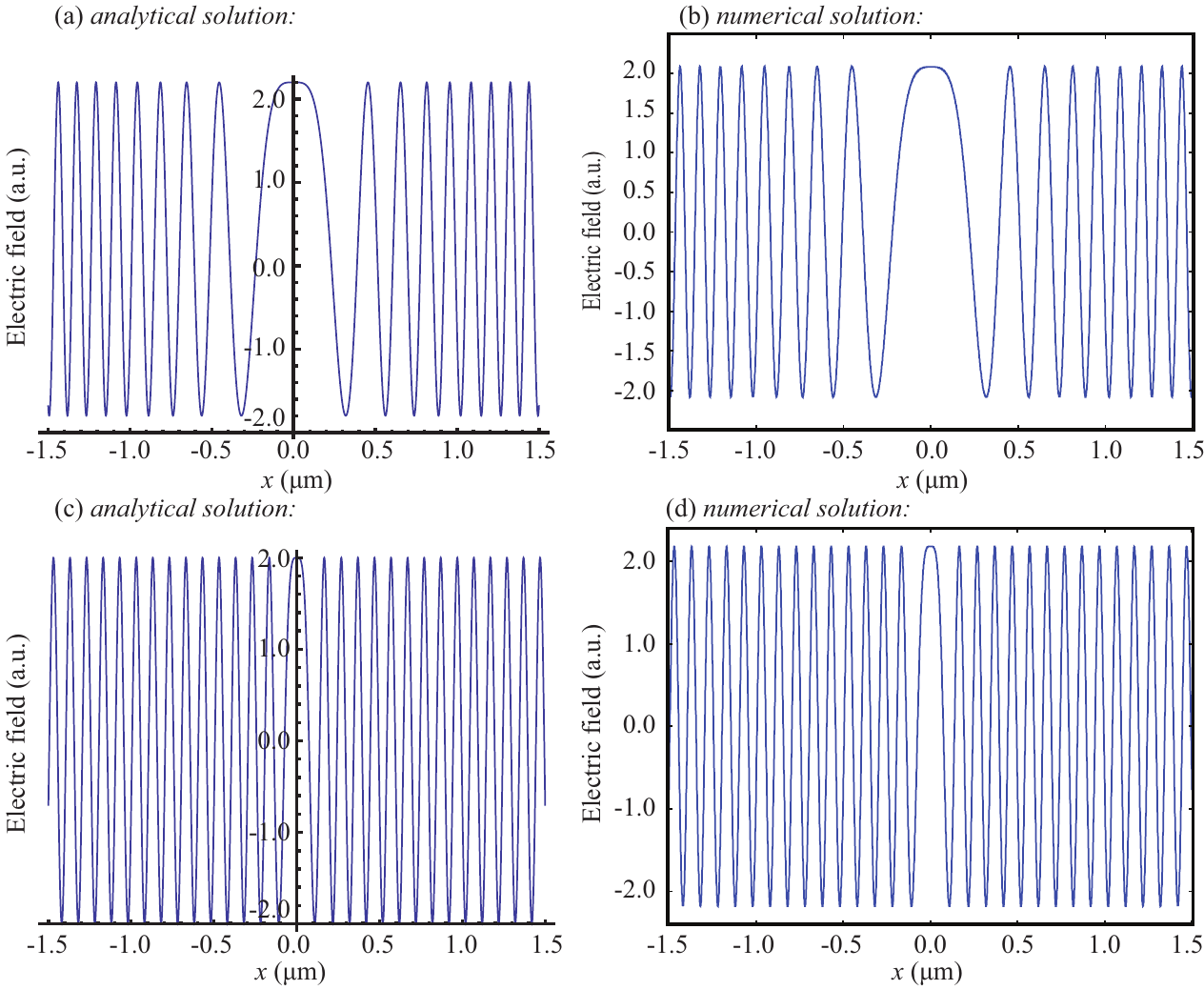}
  \caption{Comparison of the analytical and numerical results for the electric field. We plot $E(x)$ at $t=0$. (a)~Analytical solution for $\rho = \unit{10}{\micro\meter}$. (b)~Numerical solution for $\rho = \unit{10}{\micro\meter}$. (c)~Analytical solution for $\rho = \unit{1}{\micro\meter}$. (b)~Numerical solution for $\rho = \unit{1}{\micro\meter}$.}
  \label{fig:Figure3}
\end{figure}

In order to validate our exact analytical solution [Eq.~(12)], we compare the waveforms with results obtained from a direct simulation of Maxwell's equations. We use a finite element method (COMSOL Multiphysics) to discretize Maxwell's equations, using perfectly matched layers at the left and the right of the structure to close the simulation domain. For the simulation results shown in Figs.~\ref{fig:Figure3}(b) and~(d), we have used the following parameters: $\lambda_0 = \unit{1}{\micro\meter}$ and $\epsilon_\mathrm{eff}(\lambda_0) = \mu_\mathrm{eff}(\lambda_0) = 1$. Figures~\ref{fig:Figure3}(a)-(b) and Figs.~\ref{fig:Figure3}(c)-(d) are for different transition steepness. We see that there is excellent agreement between the analytical and numerical results.

\section{Conclusion}
We have investigated electromagnetic wave propagation through a graded index interface between a right-handed and a left-handed material. We derived an exact analytic wave solution when the index of refraction varies according to a hyperbolic tangent function. This solution is valid for arbitrary steepness of the index transition, even when the effective constitutive parameters vary on the scale of the vacuum wavelength where the traditional approximate methods (e.g., SVEA) cease to work. We have validated our analytical solutions by accurate numerical simulations using a finite element method. Our analytical model allows for arbitrary dispersion.

\section*{Acknowledgments}
P.~T.\ is a Research Assistant of the FWO-Vlaanderen. Work at the Vrije Universiteit Brussel was financially supported by the \textsc{FWO}-Vlaanderen, by the Belgian Science Policy Office (grant no.\ IAP-VI/10: Photonics@be), and by the Research Council (\textsc{OZR}) of the university.

\end{document}